# Resistive switching in $HfO_{2-x}/La_{0.67}Sr_{0.33}MnO_3$ heterostructure: An intriguing case of low H-field susceptibility of an E-field controlled active interface

*Vivek Antad,*[1,2,#], *Parvez A. Shaikh,*[3,#] *Abhijit Biswas,*[4,#] *Shatruhan Rajput,*[1] *Shrinivas Deo,*[5] *ManjushaShelke,*[6] *Shivaprasad Patil,*[1,*] *and Satishchandra Ogale*[1,7,*]

[1]Department of Physics and Centre for Energy Science, Indian Institute of Science Education and Research (IISER) Pune, Pune 411008, India

[2]Department of Physics, MES's Nowrosjee Wadia College of Arts and Science, Pune 411001, India

[3]Department of Physics, Y & M AKI's Poona College of Arts, Science and Commerce, Pune 411001, India

[4]Department of Materials Science and Nanoengineering, Rice University, Houston, Texas 77005, USA

[5]Centre for Materials Characterization, CSIR-NCL, Pune 411008, India

[6]Physical and Materials Chemistry Division, Polymer and Advanced Materials Laboratory, CSIR-NCL, Pune 411008, India

[7]Research Institute for Sustainable Energy (RISE), TCG Centres for Research and Education in Science and Technology (TCG-CREST), Kolkata 700091, India

ABSTRACT:

High-performance non-volatile resistive random access memories (ReRAM) and their small stimuli control are of immense interest for high-speed computation and big-data processing in the emergent Internet of Things (IOT) arena. Here, we examine the resistive switching (RS) behavior in growth controlled $HfO_2/La_{0.67}Sr_{0.33}MnO_3$ heterostructures and their tunability under low magnetic field. It is demonstrated that oxygen-deficient $HfO_2$ films show bipolar switching with high on/off ratio, stable retention, as well as good endurance owing to the orthorhombic-rich phase constitution and charge (de)-trapping-enabled Schottky-type conduction. Most importantly, we have demonstrated that, the RS can be tuned by a *very low* externally applied magnetic field (~0-30 mT). Remarkably, application of a magnetic field of 30 mT causes the RS to be fully quenched and frozen in the high resistance state (HRS) even after the removal of magnetic field. However, the quenched state could be resurrected by applying higher bias voltage than the one for initial switching. This is argued to be a consequence of the electronically and ionically "active" nature of the $HfO_{2-x}$/LSMO interface on both sides, and its susceptibility to the electric and low magnetic field effects. This result could pave the way for new designs of interface engineered high-performance oxitronic ReRAM devices.

INTRODUCTION:

The emergent gigantic increase in the digital data processing needs (*e.g.* IOT) and big-data workloads (e.g. machine learning) are posing immense challenges to the current digital computing strengths, speeds and storage densities.[1–3] Therefore, new memory device concepts and architectures capable of rendering significantly higher storage density, high throughput, energy-efficiency, and area-efficient information processing are required for data-centric computation[4]. In this regard, resistive switching materials are rapidly emerging as one of the most promising candidates for non-volatile resistive random access memory (ReRAM) applications.[5,6] ReRAM, generally a metal-insulator-metal (MIM) device with two-terminal junction, embodies the resistive switching (RS) phenomenon at its core wherein the device resistance can be locally altered between two stable states, i.e. a low resistive state (LRS) and a high resistive state (HRS) in response to the applied electric bias. This device is also termed as a memristor. Although numerous materials have been explored for this beguiling RS application over the years, however research on the novel materials, device architectures, and interface engineered functional metal oxide systems is still being very actively pursued for realizing enhanced performance and robust modulation.[7,8] Among various RAM alternatives, ReRAM stands out for its simple structure,[9] wide material range,[10] versatility of fabrication,[11–13] high endurance and retention,[14] and high-speed operation.[15] Even though the last decade has witnessed explorations of a variety of dielectric materials for the ReRAM applications, presently the research on multioxide-interface heterostructures[16,17] is on the rise due to the astonishing changes in RS properties encountered at the corresponding interfaces[18]. Such interface-engineered oxide heterostructures render stable alterations in LRS-HRS conditions,[10] modify conduction mechanisms,[19] improvise memory window enlargements,[20] and elevate the overall RS performance[21] thereby proving to be highly suitable for the efficient as well as reliable high-performance memory applications.

Herein, we have examined the RS phenomena in ReRAM devices comprised of heterostructure of $HfO_{2-x}$ and $La_{0.67}Sr_{0.33}MnO_3$ (LSMO) grown on $LaAlO_3$ (LAO) substrate using pulsed laser deposition (PLD). The intent was to investigate the intriguing interface between LSMO, which is capable of a concurrent change in its magnetic and electronic properties upon the application of low magnetic field[22] and $HfO_2$, a high-κ-dielectric with potentially interesting properties emanating from near-interface defects and trap-state enabled polaronic screening effects.[23] Indeed, LSMO is demonstrated to show electric field induced resistance (EIR) change, which also changes the magnetic state of the system.[24] Therefore, we conjectured that such an interface could bear a parameter space that has hitherto not been examined for the RS phenomenon and could potentially render the desirable electric field effect induced low magnetic field tunability of the same. As reported herein, this is indeed the case. Due to high-κ dielectric character and other interesting properties, there have been several studies of the RS phenomenon in the case of $HfO_2$ thin films.[25–29] These have involved various growth methods for the $HfO_2$ films as well as use of different end-electrode materials. The general physical picture that has emerged with good supporting evidence is one of filament mechanism and the pinching off and on of the filament paths by oxygen vacancy related traps under the applied field cycles.

Our current-voltage (*I-V*) analysis reveals a strong RS response of ~$10^3$ to ~$10^5$ depending on $O_2$ partial pressure (100 mTorr/10 mTorr/1 mTorr) maintained during the growth of $HfO_2$ (for incorporating oxygen vacancies). The devices grown at lower partial pressures are found to show much better retention and endurance performances, indebted to orthorhombic-rich phase and trapping-de-trapping-enabled Schottky-type conduction mechanisms. Moreover, we observe a high positive compliance voltage-enabled tailoring of the set voltage $V_{set}$, from -8.5 V to -2.5 V. Most importantly, we have demonstrated that, the RS can be tuned by a *very low* externally applied magnetic field (~0-30 mT). Remarkably, the RS is found to be quenched at a critical magnetic field of 30 mT and frozen in HRS even after the removal of the applied magnetic field. Moreover, we further demonstrate that the original RS state can be recovered by applying a considerably higher $V_{set}$ (of about -14 V) as compared to -8.5 V (i.e. the $V_{set}$ observed sans magnetic field).

EXPERIMENTAL SECTION:

**Device Fabrication.** Thin films of $HfO_2$ and LSMO were grown on $LaAlO_3$ (001) substrates by PLD (248 nm KrF excimer laser, repetition rate 5 Hz and energy density ~3 J/cm$^2$). The growth temperature was maintained at 700°C throughout the depositions. The target-to-substrate distance was kept ~40 mm. After growth of ~100 nm LSMO, the film was partially masked and a fresh ~20 nm LSMO over-layer was grown, followed by ~15 nm $HfO_2$ to avoid contaminations at the key oxide-oxide interface. $O_2$ partial pressure was always kept at 100 mTorr during the LSMO depositions (for buffer as well as the over-layer). After the growth, LSMO films were post-annealed at the same growth temperature and at 500 mTorr oxygen pressure for 30 min. For the $HfO_2$ growth, all the growth parameters were the same as for LSMO; however the oxygen pressure was deliberately changed in different depositions *viz.* 100 mTorr, 10 mTorr and 1 mTorr to create a controlled variation of oxygen vacancy concentrations in the deposited film.

**XRD.** The crytalline nature of the films and their phases were studied with XRD characterizations performed on Bruker powder X-ray diffractometer (D8 Advance).

**XPS.** The X-ray photoelectron spectroscopy characterizations of $HfO_2$/LSMO devices (with $HfO_2$ grown at 10 mTorr) done with Thermo Scientific K-Alpha$^+$ spectrometer.

**AFM experiments.** AFM was done with JPK Nanowizard II (Germany) for characterizing roughness and thickness of involved thin films. Micro-fabricated commercially available cantilevers from MicroMasch, with a typical force contact ~0.05 N/m, were used. All AFM characterizations were done in contact mode.

***I-V* measurements.** Keithley 4200 SCS source meter, reinforced with a state-of-the-art probe station (Semiprobe Systems, USA), was used for *I-V* characterizations. To avoid the pin-hole related issues (if any), rather than creating metal contact patterns, the tungsten (W) probe tips of 7 μm tip diameter, were directly used on LSMO buffer layer (bottom electrode) as well as on $HfO_2$ (top electrode). Thanks to the super fine X-Y-Z motion adjustments of probe-arms and probe tip holders, the mechanical pressing or physical penetration is avoided while creating a delicate but trusted contact between the tip and film surfaces. The bottom electrode was grounded whilst a variable potential was applied at the top electrode. A systematic dual voltage sweep cycle +10V → 0V → -10V → 0V → +10V was employed (indicated as ±10V dual cycle, unless otherwise mentioned) with 0.05V voltage step having a constant sweep delay of 0.7s (unless otherwise mentioned) and the resultant electrical current was collected (measured) in the current-perpendicular-to-plane (CPP) mode.

**Electromagnetic measurements.** A Keithley 2400-C source-meter was employed to send the accurately measured currents through the indigenously built electromagnet and subsequently generated magnetic field was measured with an InAs Hall Probe using Digital Gaussmeter from SES Instruments, Roorkee (Model: DGM-102).

RESULTS AND DISCUSSION:

**Thin film growth, structural and morphological characterizations.** Thin films of $HfO_2$ and $La_{0.67}Sr_{0.33}MnO_3$ (LSMO) were grown on $LaAlO_3$ (001) substrate by pulsed laser deposition (PLD). Oxygen partial pressure was kept at 100 mTorr during the LSMO film growth, whereas it was deliberately varied to 100 mTorr, 10 mTorr and 1 mTorr during the $HfO_2$ deposition to create oxygen deficiencies. As shown in Figure 1a, ~100 nm LSMO film acts as the bottom electrode while ~15 nm $HfO_2$ over-layer acts as a dielectric layer. Tungsten (W) probes were directly employed as the metal contacts. The AFM topography images of PLD grown $HfO_2$ and LSMO film surfacesare shown in Figure 1b,c, demonstrating the smooth surfaces of deposited films with the surface roughness of less than 1 nm, thus indicating the high-quality of growth.

The XRD patterns show crystalline nature of $HfO_2$/LSMO heterostructures (Figure 1d) where the influence of $O_2$ partial pressure on the crystalline nature of $HfO_2$ is distinctly noticeable. A mixed-phase (monoclinic and orthorhombic), with predominant monoclinic phase as well as the ferroelectricity-laden polar orthorhombic phase is evident for $HfO_2$ deposited at 100 mTorr $O_2$ partial pressure. For films deposited at lower partial pressures, such as 10 mTorr and 1 mTorr during the $HfO_2$ growth, the increase in the orthorhombic/monoclinic ratio, as the dominance of monoclinic phase in $HfO_2$ is seen to be significantly reduced, giving rise to a orthorhombic-rich crystalline phase.[30,31] As is borne out in this study, and also noted by others, RS in a single-phase material is always known to render a stable and reliable performance in the case of oxide-based ReRAM device.[32–37]

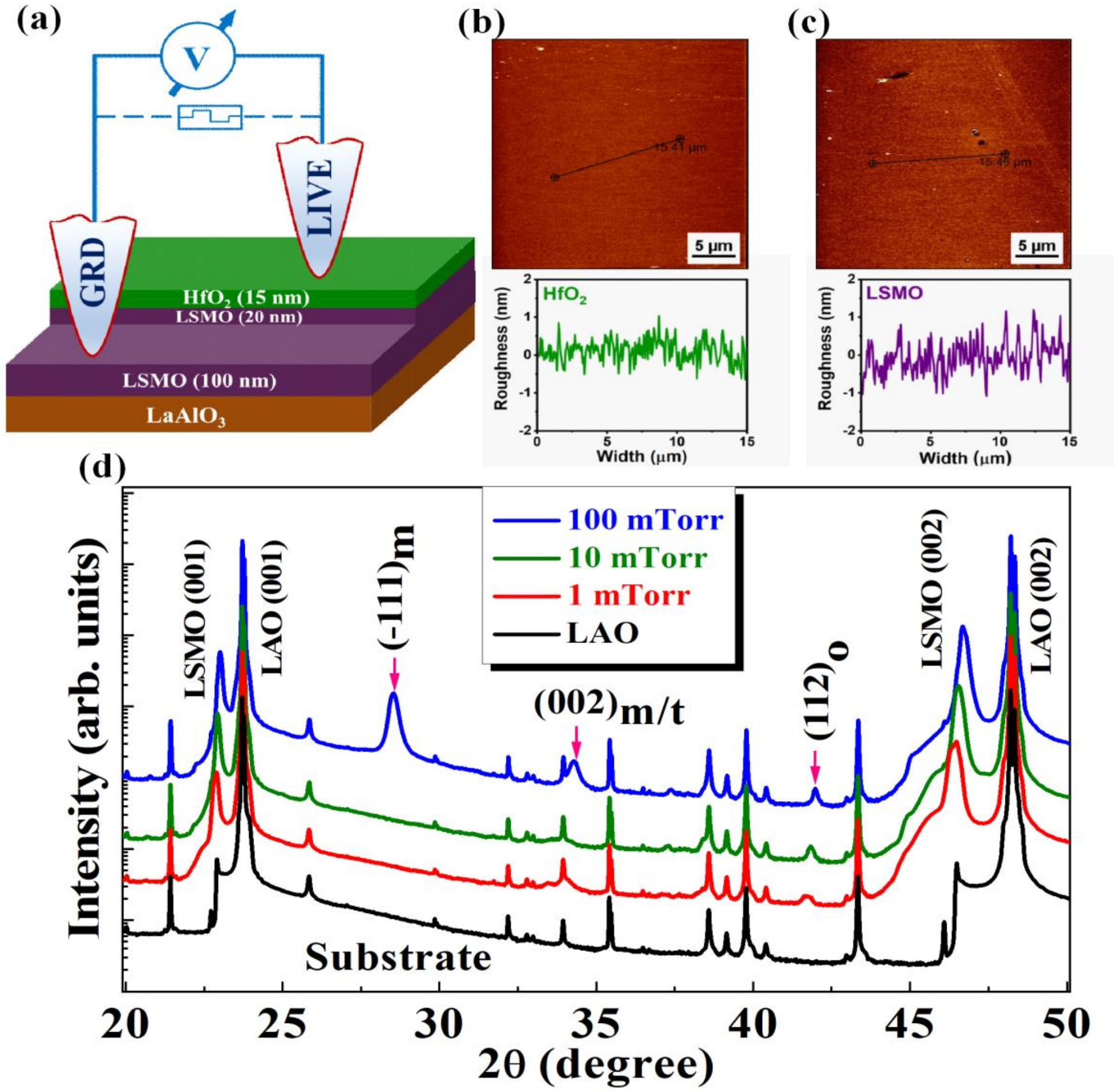

**Figure 1.** The device architecture and $HfO_2$/LSMO/LAO heterostructure characterizations. (a) Schematic of the resistive switching device (memristor) architecture in the current perpendicular to plane (CPP) mode. The bottom electrode (on LSMO) was grounded (GRD) whereas the bias was applied to the top electrode (on $HfO_2$), thus referring it as LIVE. Typical AFM topography images and respective roughness profiles for: (b) $HfO_2$ (grown at 10 mTorr) deposited on (c) LSMO (grown at 100 mTorr). (d) XRD patterns of $HfO_2$/LSMO grown on LAO substrate at various $O_2$ partial pressures (100 mTorr/10 mTorr/1 mTorr) obtained by using a powder X-ray

diffractometer. The $O_2$ pressure was varied only during the growth of $HfO_2$, whereas for LSMO it was always kept fixed at 100 mTorr.

**$HfO_2$ growth pressure dependence of resistive switching.** Figure 2a shows distinctive semi-log *I-V* plots for $HfO_2$/LSMO/LAO heterostructures, reflecting a clear influence of the $HfO_2$ oxygen growth pressure dependence on the RS performance.

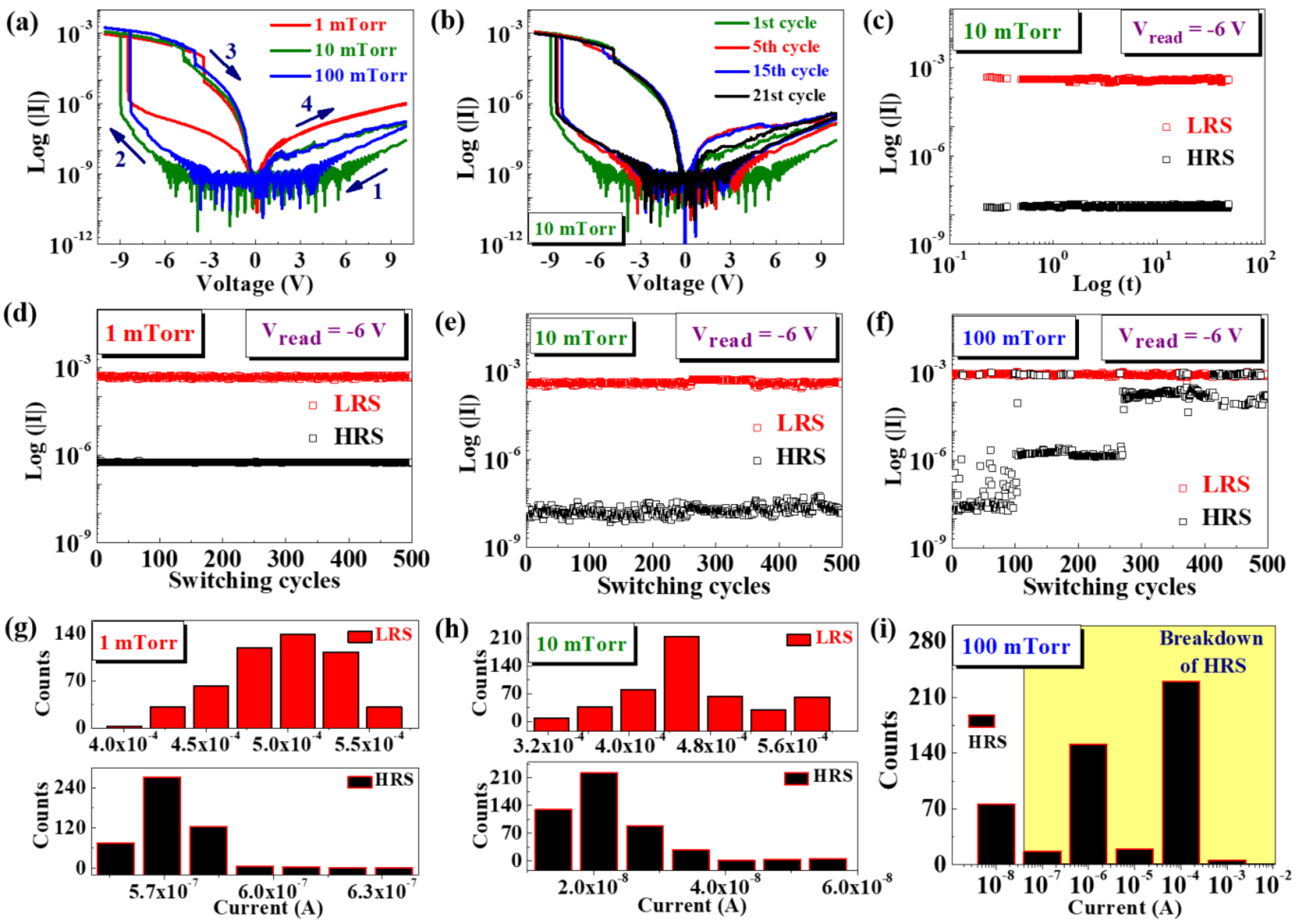

**Figure 2.** Influence of $O_2$ partial pressure on RS of $HfO_2$/LSMO/LAO devices. (a) Semi-log *I-V* plots for three devices with $HfO_2$ grown at different $O_2$ partial pressures. Arrows depict the sequence of dual voltage sweep. A case of 10 mTorr device depicting: (b) stability of RS through multiple dual voltage sweep cycles and c) retention studied at $V_{read}$= -6V. Repetitive endurance analysis at $V_{read}$= -6V: (d) 1 mTorr case, e) 10 mTorr case, and (f) 100 mTorr case along with their respective current histograms shown in (g), (h), and (i), respectively.

Primarily, for each device, independent of the partial pressure, when subjected to a dual voltage sweep cycle ±10V (+10V → 0V → -10V → 0V → +10V), a clear switching from HRS to LRS is noted as the applied reverse bias voltage at the top electrode ($W/HfO_2$ junction) reaches around -8.5 V, causing the device to go into set mode ($V_{set}$). However, as can be seen, exactly opposite to $V_{set}$, the reset processes do not take place at some specific voltage ($V_{reset}$), but need a range of positive voltage values to bring the device back to its original HRS conditions. This unique nature of different SET/RESET voltage could be attributed to the presence of non-lattice oxygen anions in $HfO_2$ incorporated during growth, which could potentially migrate to and fro the electrodes under bias.[29]

Interestingly, the switching ratio ($R_{HRS}/R_{LRS}$) for the devices wherein the $HfO_2$ layer is grown at 100 mTorr and 10 mTorr is higher (~$10^5$) as compared to the 1 mTorr case (switching order is ~$10^3$). Furthermore, without the need of any compliance current, the resistive switching reliability is visible, especially for 10 mTorr case, through the repetitive voltage sweep cycles (Figure 2b), retention (Figure 2c) and endurance (Figure 2e) plots. However, a comparison of repetitive endurance at $V_{read}$ = -6V (Figure 2d,e,f) for each device reveals that, even though for the 100 mTorr case the switching is ~$10^5$, it becomes critically unstable after a few endurance cycles as compared to the devices where $HfO_2$ is grown at lower $O_2$ partial pressures (10 mTorrand 1 mTorr). The histograms of current values plotted in Figure 2g,h also hint towards the switching stability. At $V_{read}$ = 6V, for 1 mTorr case, the average HRS and LRS values are found around $1.1\times10^7\ \Omega$ and $1.1\times10^4\ \Omega$, respectively, thus showing an average switching order of about $10^3$. In contrast, for 10 mTorr case, HRS and LRS values obtained are $0.3\times10^9\ \Omega$ and $1 \times 10^4\ \Omega$, respectively, suggesting an average switching ratio of the order of about $10^5$. However, as seen in Figure 2i, the breakdown of HRS values clearly suggests the instability for 100 mTorr case, even though the initial switching order is ~$10^5$.

Such instability in RS for the 100 mTorr case could be attributed to its mixed phase nature (Figure 1d); an oxygen-rich environment during $HfO_2$ growth evidently coercing it to grow into a mixed phase state made up of monoclinic and orthorhombic phases. Lee *et al.* witnessed similar instability of RS for oxygen-rich $ZrO_2$ ReRAM device through the repetitive endurance cycles.[36]However, upon effectively improving the crystalline nature (from mixed to single phase) of $ZrO_2$ film by doping tetravalent cesium (Ce), stable RS performance was achieved. Kim *et al.* also showed that the single crystallinity of oxides ultimately leads to stable and long-term RS characteristics.[38] An oxygen-rich environment may not necessarily always be ideal for the oxide thin films to generate single-phase films with a stable and reliable RS performance. Interestingly, presence of monoclinic-rich phase within $HfO_2$-based oxide devices, is known to degrade the physical properties *viz.* ferroelectricity, dielectric constant.[37,39] Thus in our case it is possible that the instability of RS and the endurance degradations contemporaneous in 100 mTorr case may be due to the presence of mixed phase (with dominant monoclinic phase along with the presence of orthorhombic phase).[40] Moreover, an oxygen-rich RS device is known to face gas evolution problems where the negatively charged oxygen ions form $O_2$ gas that can accumulate over the time, thus making certain parts of the device more metallic in nature along the grain boundaries,[35] converting HRS permanently into LRS,[41] as is evident in our 100 mTorr case. Hence, for stable RS operations, controlled optimum density of oxygen defects in the oxide ReRAM device seems absolutely necessary.[38]Such controlled oxygen vacancies within an oxide film also sensitively alters its crystalline nature, thus evidently affecting the RS parameters of that oxide device.[34,35] As clearly seen in Fig. 1d, when $O_2$ partial pressure is low (10 mTorr or 1 mTorr), due to the induced oxygen vacancies, the crystalline nature of $HfO_2$ gets dominated by orthorhombic phase over the monoclinic one and this change of crystalline nature gives rise to a stable RS as can be observed through the repetitive endurance cycles, presented in Figure 2d,e. Hence, it is viable to say that the oxides in ReRAM with single-phase crystalline nature can make RS more stable and when they are grown with the different density of oxygen vacancies, produce dependable retention and endurance cycles.[42]

Among three devices compared in Figure 2d,e,f, from the stability and reliability standpoint, 1 mTorr and 10 mTorr devices perform much better than the 100 mTorr one. While, considering the stability and higher switching order, 10 mTorr $HfO_2$/LSMO device emerges to be the better option. Hence, we investigated it further to tune the RS behavior through various parameters. Henceforth, in the main text, we will only discuss about the $HfO_2$/LSMO device, in which $HfO_2$ was grown at 10 mTorr $O_2$ partial pressure, whereas the results for the other two-cases are shown in the supplementary sections.

**Measurement parameter dependence of resistive switching.** The RS is known to be sensitively linked not only to the physical parameters such as film thickness,[16,43] but also to the applied current-voltage values[44] set during the *I-V* measurements. In this regard, Figure 3 shows the tuning of $V_{set}$ in $HfO_2$/LSMO devices by altering the intrinsic measurement parameters, such as voltage sweep speed delay (Figure 3a and supplementary material Figure S1) and the positive compliance voltage (Figure 3b and supplementary material Figure S2). As seen in Figure 3a, $HfO_2$/LSMO device was subjected to a dual voltage sweep cycle, albeit with intentionally induced delays during the scan *viz.* 0.7s, 1.3s, and 1.6s, changing $V_{set}$ values consecutively from about -9V, -8V and -7V, respectively. Further, as can be seen from supplementary material

Figure S1, irrespective of the $O_2$ partial pressure during $HfO_2$ deposition, the trend of $V_{set}$ increment in $HfO_2$/LSMO devices is quite similar with increasing scan delay, without altering the LRS current. Similarly, the influence of positive compliance voltage on $V_{set}$ is shown in Figure 3b. Here, upon the application of a positive compliance voltage of +10V at the W/$HfO_2$ junction during the applied dual cycle, $V_{set}$ was found to be around -9V, which then systematically increased to -2.5V as the positive compliance voltage was reduced from +10V to +2V, respectively. During these measurements, a negative bias was kept constant, -10V. Here, the linear increase in $V_{set}$ as a function of decreasing positive compliance voltage for $HfO_2$/LSMO devices can be seen in supplementary material Figure S2 and is independent of $O_2$ partial pressure at which $HfO_2$ is grown. Strikingly, in absence of compliance voltage no RS is observed, indicating that a minimum positive compliance voltage is necessary to take the device from its HRS to LRS condition through the resistive switching mechanism, and generate a certain minimum order of on/off ratio.

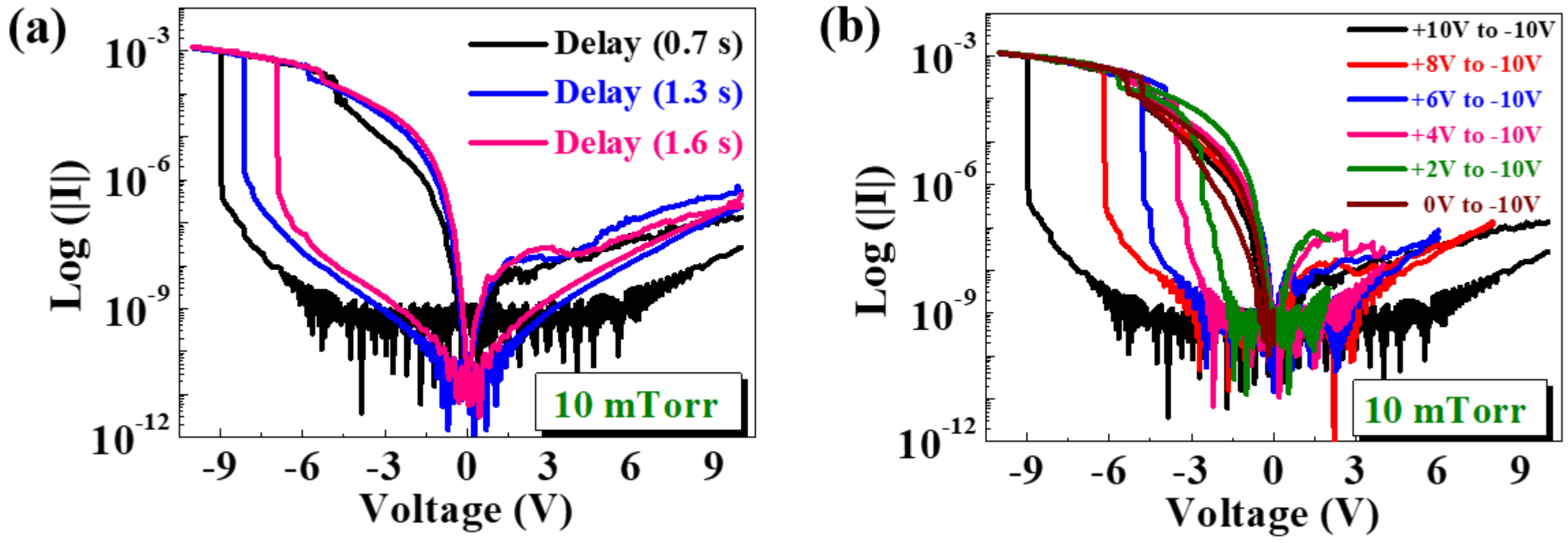

**Figure 3.** Influence of measurement parameters on RS of $HfO_2$/LSMO/LAO devices with $HfO_2$ grown at 10 mTorr $O_2$ partial pressure. (a) Semi-log *I-V* plots for RS studied as a function of induced voltage sweep delay (0.7s, 1.3s, and 1.6s). (b) Semi-log *I-V* plots depicting the tuning of $V_{set}$ by applied positive compliance voltage.

**Salient *low* magnetic field modulation of resistive switching.** Apart from the intrinsic parameters such film thicknesses and oxygen vacancies, extrinsic parameters *viz.* thermal energy, electromagnetic radiation, and external magnetic field can offer potential controls for RS in ReRAM devices that can influence remotely, thereby adding the additional degrees of freedom for RS manipulation.[16,45–47] Among these, externally applied magnetic field can be a true remote control for RS modulations as it can be used in non-intrusive as well as non-destructive ways.[16] The details of such research work in the literature are compared with the current resultsin Table S1 of Supporting Information. A colossal magnetoresistance manganite LSMO electrode (showing EIR changes, which also changes the magnetism of the system[24]) was precisely used for this purposein the RS architecture acting as a highly conducting Zener double exchange ferromagnet under the small magnetic field, and as shown, it indeed delivered on this front.

Supplementary material Figure S3 demonstrates the schematics of $HfO_2$/LSMO/LAO device kept in the external magnetic field during *I-V* measurements and Figure 4 summarizes the influence of *very low* magnetic field (0-30 mT) on the RS parameters of the device. As seen in Figure 4a, in comparison with the case of absence of magnetic field, application of a mere 10 mT magnetic field distinctly changes $V_{set}$ from around -9V to -8V, accompanied with a dramatic change in the switching order from ~$10^5$ to ~$10^3$. Interestingly, the HRS of the device, during voltage cycle, is itself found to be elevated due to the application of magnetic field than the actual HRS sans magnetic field. Surprisingly, a magnetic field of 30 mT is seen to cause quenching of the RS, as evident from Figure 4a. Along the same line, retention comparisons of LRS (Figure 4b) at $V_{read}$ = -6V also suggest a gradual minimization of RS window due to the successive increment in the externally applied magnetic field strength.

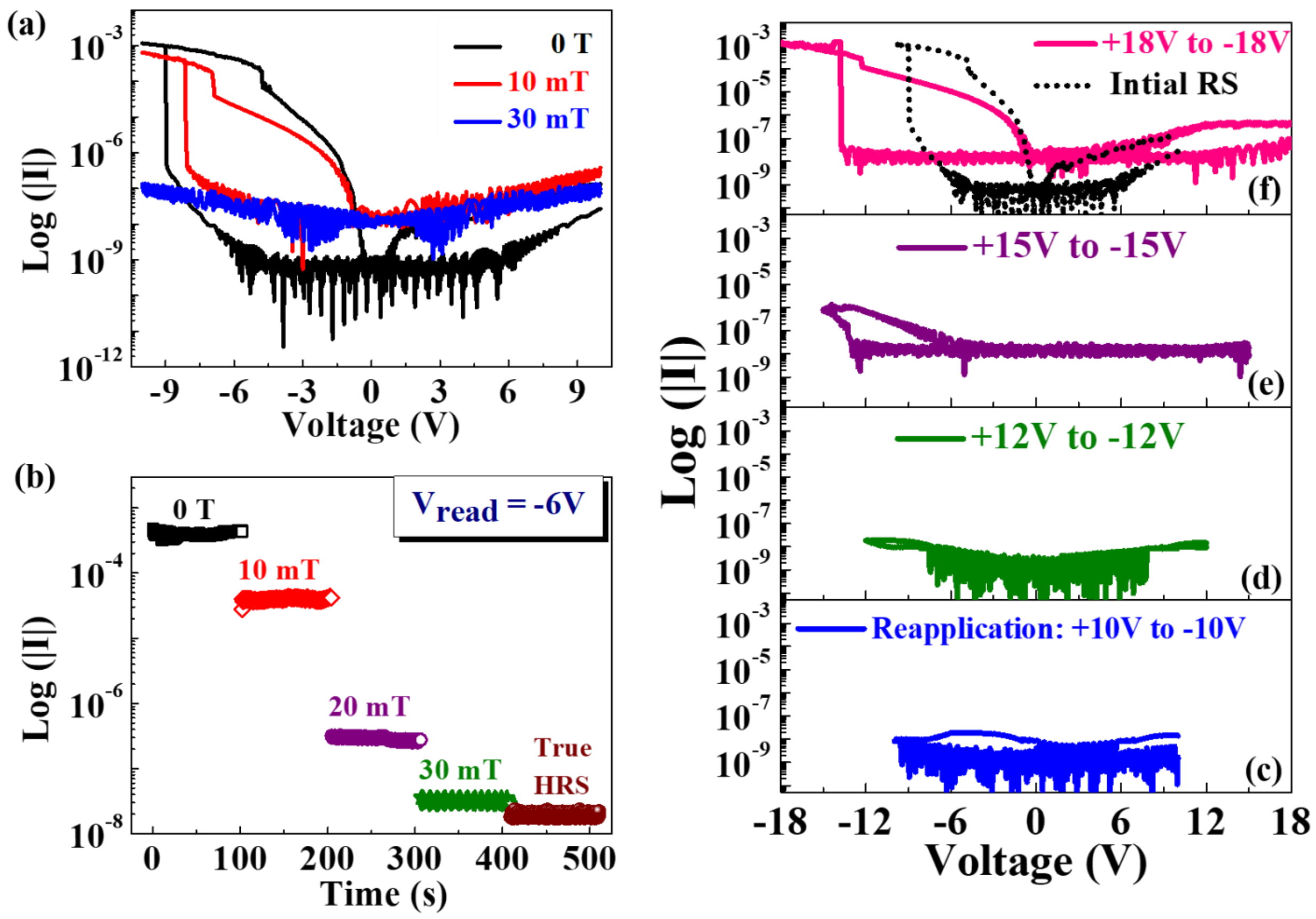

**Figure 4.** Influence of magnetic field on RS of $HfO_2$/LSMO/LAO devices with $HfO_2$ grown at 10 mTorr. (a) Semi-log *I-V* plots, without and with externally applied magnetic field, show the quenching of RS for critical magnetic field of 30 mT. (b) Magnetic field dependent retention analysis at $V_{read}$= -6V, compared with HRS state of the device sans magnetic field (True HRS). The sequential reappearance of RS by subjecting the device to linearly increased dual voltage sweep: (c) ±10V, (d) ±12V, (e) ±15V, and (f) ±18V, respectively. The reappearance of RS at ±18V (pink curve) is compared with the original RS for the same device (black dotted curve) before the application of external magnetic field.

Furthermore, at magnetic field strength of 30 mT, due to the complete quenching of RS, the HRS of the device nearly approaches the 'true HRS' which is acquired without magnetic field.

Ubiquitously, it was found that at any applied magnetic field, modifications in the RS parameters (such as the switching order and $V_{set}$) are irreversible even after the removal of magnetic field, implying the persistent influence of the concurrent magnetic field induced changes in the electrical conduction of LSMO at the $HfO_2$/LSMO interface and related RS.

Interestingly, resurrecting the RS effects in the device, which were quenched by applying *only* 30 mT magnetic field strength, is not that straight forward. As shown in Figure 4c, after removing the magnetic field, re-application of the same ±10V dual voltage sweep cycle fails to resurrect the quenched RS of the device. A further sequential increment in the applied voltage (±12V) too, could not cause the device to switch from HRS to LRS (Figure 4d). However, the regeneration of RS switching behavior is realized when the dual voltage sweep cycle of (±15V) is applied to the device (Figure 4e), altering the HRS to LRS at $V_{set}$ around -12V with RS order of $\sim 10^3$. A further advancement in the applied voltage dual cycle of (±18V) strikingly returns the device back to the original LRS state (Figure 4f) along with the original switching order of $\sim 10^5$. Moreover, as can be seen, $V_{set}$is now increased to -14V (in case of applied ±18V) from its original value of -8.5V (without any magnetic field as shown in Figure 2a).

**The conduction mechanism.** Considering the band gap, interface effect, presence of oxygen vacancies, applied voltage sweep, and top/bottom electrode materials, an interesting and peculiar mechanism is clearly operative in our ReRAM device configuration. At room temperature LSMO is known to be a narrow band gap ($E_g$) material (~0.60 eV)[48] and $HfO_2$ is a wide bandgap material (~5.80 eV).[49–51] The precise band alignment at the interface is guided by the relative positions of the Fermi energy level ($E_F$) depending on the oxygen defects introduced during growth.[48] Furthermore, it is known that $HfO_2$ grown under oxygen deficient environment shows

*p*-type conductivity,[49–51] while moderate *p*-type behavior is also suggested for LSMO.[52,53] This essentially renders a *p-p* type oxide-oxide heterojunction,[54] albeit with differing carrier densities on both sides of the interface. The latter should render a degree of charge redistribution and therefore band bending at the interface. Additionally, the observed strong rectification in *I-V* (Figure 2a) for the present $HfO_2$/LSMO ReRAM system, hints towards more of a MIM type hetero-junction. A linear region for $\ln(I)$ vs $\sqrt{V}$, as shown in supplementary material Figure S4, explicitly reveals that the Schottky emission mechanism is responsible for the observed *I-V* curves, confirming the formation of the Schottky junction.[55]

In the case of oxide heterojunction interfaces, the most common conduction mechanism known for switching from HRS condition to LRS condition (and vice-versa) is the formation (or disruption) of the conductive filaments (CF).[10] As discussed in several references on RS in $HfO_2$films,[25–29] the same CF principle seems to hold in the case of this material system as inferred from multiple experimental evidences. In this mechanism, once the filament is induced, its repeated clipping and reconnection can be realized by controlled reverse and forward electric field application. When both sides of the active oxide element are normal metals the actions occur within the oxide element, unless conditions exist for metal ion in-diffusion or oxidation of metal by oxygen out-diffusion. In the case of our device, the situation is interestingly different because the interface between $HfO_2$ and LSMO is an "active one" with electronic as well as ionic (oxygen) dynamics available on both sides. Thus, oxygen vacancies (including their E-field driven movements) and related trap states can be operative on the $HfO_2$ side, while E-field induced modification of the $Mn^{3+}/Mn^{4+}$ ratio along with possible oxygen vacancy movement are operative on the LSMO side of the interface. With the presence of the LSMO "active layer" on one side of the interface the device design provides for the electric field effect in manganite to operate in tandem with the RS effect in the $HfO_2$ layer.[56–58] The concurrent susceptibility of magnetic and electronic state in LSMO (via double exchange) to the applied electric and magnetic field adds further intriguing and interesting effects involving low magnetic field tunability. It should be noted that the Curie temperature for $La_{0.67}Sr_{0.33}MnO_3$ ($Mn^{3+}/Mn^{4+}$ ratio = 0.67) is much above room temperature (RT). However, as per the phase diagram,[59–61] the changes in the $Mn^{3+}/Mn^{4+}$ ratio can either push the system to ferromagnetic insulating state ($Mn^{3+}$ rich side) or ferromagnetic state with a lower than room temperature RT Curie temperature or even a charge ordered insulating state ($Mn^{4+}$ rich side). As stated above, the E-field modification on the LSMO side under HRS condition can push the material to one of the two sides of the phase diagram depending on the sign of the applied field from the top electrode.

The filament formation mechanism in the case of binary oxide systems including the $HfO_2$ case has been discussed in some interesting literature reports.[62–64] The corresponding discussion of the mechanism has been primarily focused on the oxygen vacancy ($V_O$) formation energy, its charge state, mobility, and the energetics involved in the vacancy isolation *Vs* cohesion and mobility. Isolated $V_O$'s favor $2^+$ charge state and are fairly mobile in most binary oxides. In the case of $HfO_2$ the activation energy for their transport is 0.7 eV. When their cohesion occurs, they attain $1^+$ or neutral charge state and become almost immobile (activation energy for the $HfO_2$ case jumps to 3 eV), rendering stable chain-like conducting filament form. This configurational transition from isolation to cohesion state requires nucleation and is clearly a many particle process. It has been suggested that oxygen deficient $HfO_{2-x}$ (with x in the range 0.25 to 0.5) is prone to the filament nucleation. It must also be mentioned that the charge state for isolated $V_O$ being $2^+$ while that of $V_O$ in the filament state being $1^+$ or neutral, charge transfer from the electrodes is an essential concurrent process for nucleation and therefore it basically occurs near an electrode. Within the framework of this scenario, as depicted in Figure 5, we could understand the results reported herein as follows.

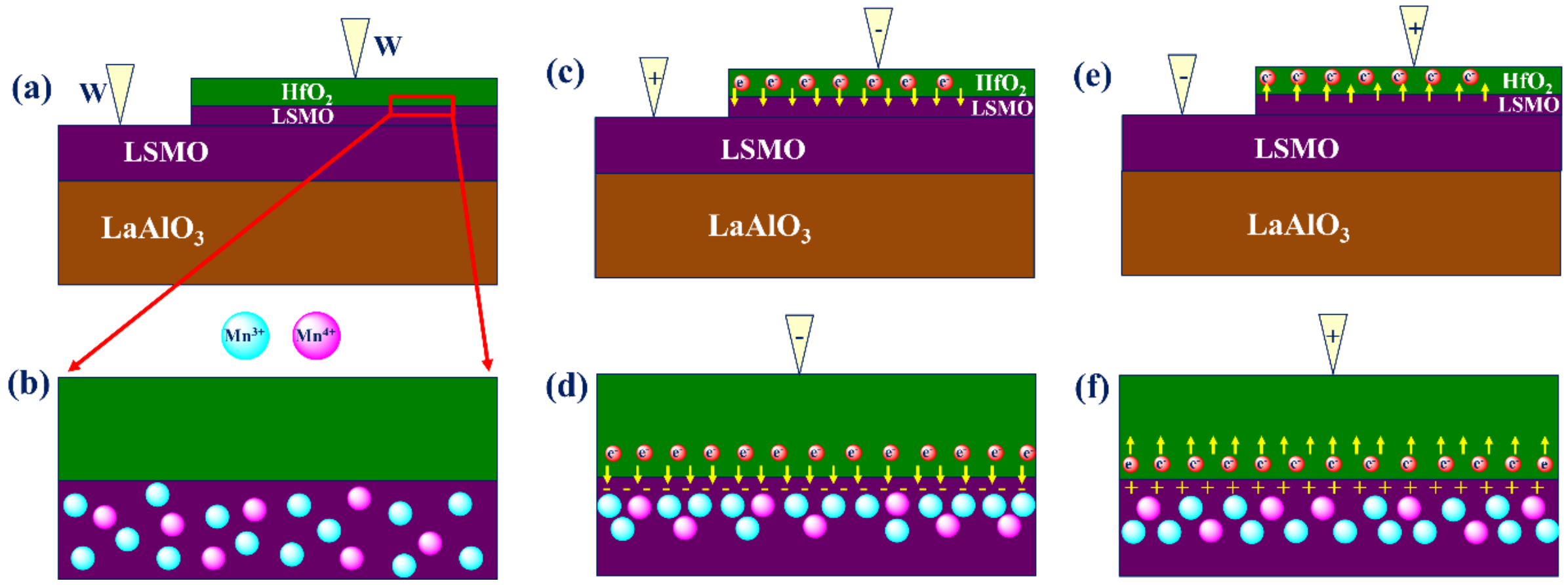

**Figure 5.** Schematic representations of the possible conduction mechanism of W/$HfO_2$/LSMO heterostructure: (a) and (b) Ideal device without applying any bias voltage, with the presence of both $Mn^{3+}$ and $Mn^{4+}$ cations in LSMO. (c) and (d) LRS condition where negative bias is applied at W/$HfO_2$. (e) and (f) HRS condition where positive bias is applied at W/$HfO_2$ junction, in the presence of applied magnetic field.

When a positive voltage is applied to the top metal point contact electrode, the $V_O$'s bearing $2^+$ charge state would get pushed towards the bottom electrode, accumulating at the interface between $HfO_2$ and LSMO. This will certainly be a fraction of all the $V_O$'s present in the layer as controlled by growth ambient pressure, because with progressive accumulation there will be a screening effect to resist further accumulation. This would essentially create an $HfO_{2-x}$/LSMO interface and the charge of the vacancies in the accumulation will be reduced by the electron injection from LSMO. Inside such an interface dressed with vacancy clusters, the valence state would be close to $1^+$ or zero, and the clusters would also be relatively stable and immobile. When the polarity is reversed and negative bias is progressively enhanced on the top tip electrode, an attractive electromotive force will be experienced by these clusters and condition will be set for the formation of the filament by the disruption of the clusters as well as cooperative exodus of other $2^+$ charged $V_O$'s. The filament will then strike at certain negative bias with the electrons injected from the tip causing partial charge neutralization from $2^+$ to $1^+$ or zero to stabilize it. This would represent the LRS state. As we reduce the negative bias, at some point the transition to the HRS state would occur, as discussed in the literature,[10,65,66] via clipping of one end of the filament near the interface, once again by local cohesion to isolation transition.

The low magnetic field tunability is equally intriguing and highly application-worthy. The most natural and reasonable explanation for the same must relate to the high sensitivity of the magnetic and electric phase diagram of manganite (LSMO in the present case) to the $Mn^{3+}$ and $Mn^{4+}$ ion concentration and its distribution in the matrix, since the $Mn^{3+,4+}$-O-$Mn^{4+,3+}$ network controls magnetism and transport (double exchange).[67,68] In the present case, the top electrode operating via the high-κ dielectric $HfO_2$ not only serves as the current (carrier) extracting/injecting contact, but also serves to modify the $Mn^{3+}$ and $Mn^{4+}$ ratio in the depletion/injection region by the electric field effect in the HRS regime(please see the scheme presented in Figure 5).

In order to gain evidentiary support for the foregoing arguments pertaining to field application induced changes in the $Mn^{3+}$ and $Mn^{4+}$ valence states, we performed x-ray photoelectron spectroscopy measurements on the fresh and E-field-subjected distinct states of the samples. Since it is a buried interface problem (10-15 nm below the top $HfO_2$ film) the task of elucidating interface valence condition is non-trivial. One approach is a destructive method wherein one sputters out the top oxide layer partially and then accesses the interface region. However sputtering of heavy element such as Hf requires high energy ions and therefore has the potential to significantly change the valence condition that one seeks to examine. Therefore we adopted an approach to do the field application and testing at the step edge of our device where the thickness of the top layer waves out progressively. We indeed found that the differences in the relative contributions of $Mn^{3+}$ and $Mn^{4+}$ between the fresh as prepared sample and the field subjected sample in a different resistive state. These data are presented in supporting information S-5(I). The consequences of destructive testing which could give erroneous results are also presented in Supporting information S-5 (II and III).

Furthering the arguments, the magnetic field can impart significant itinerancy to the electrons in the interface region which must affect the interface neutralization process of the accumulated positively charged oxygen vacancies and therefore the stability of interface clusters. This in turn would affect the electric field (voltage) controlled disruption process of the vacancy clusters and lead to change in the negative voltage on the top electrode at which filament transport would set in. Above a certain magnetic field such as 30 mT, perhaps full neutralization of vacancies takes place and hence the nucleation process for filament formation is obstructed. At a considerably higher voltage bias, field driven de-trapping of electrons may be responsible for the system to return to its initial state. The trapping-de-trapping effects on the $HfO_2$ side are clearly dependent on the type, charge state, and density of oxygen vacancy defects, which in turn should also have a significant influence of the growth ambient and conditions, as seen.

CONCLUSIONS:

In summary, we have shown a systematic and remarkable low magnetic field manipulation of RS in all-oxide $HfO_2$/LSMO/LAO heterostructures, through the control of different ‘intrinsic’ and ‘extrinsic’ parameters. Indeed, external magnetic field, as low as 10 mT, shows significant modulation in RS parameters of the device. Interestingly, a magnetic field of 30 mT takes the device away from the RS condition, settling it into the stable HRS condition even after the removal of magnetic field. Surprisingly, this HRS condition is non-reverting by the application of comparable voltage sweep cycles and its resurrection is enabled only by the application of a much higher applied voltage sweep cycles. These peculiar observations can be understood on the basis of the electronic changes imparted to the magnetic/electronic state of the manganite region near the $HfO_2$/LSMO interface through the redistribution of $Mn^{3+}$ and $Mn^{4+}$ ion concentrations. Some observations also relate to the near-interface trap states (oxygen vacancies) and the carrier trapping/de-trapping effects.

ASSOCIATED CONTENT

**Supporting Information.** The Supporting Information is available and contains: Semi-log *I-V* plots depicting influence of voltage sweep delays and positive compliance voltages for $HfO_2$-LSMO devices ($HfO_2$ being deposited at 1 mTorr and 100 mTorr $O_2$ partial pressure), $V_{set}$ variation as a function of voltage sweep delays and positive compliance voltages, schematics of LAO-LSMO-$HfO_2$ device during *I-V* measurements kept under externally applied magnetic field, plot of $\ln(I)$vs $\sqrt{V}$ suggesting possible Schottky emission mechanism, XPS analysis for freshly prepared and field subjected $HfO_2$/LSMO devices.

AUTHOR INFORMATION

**Corresponding Author**

*Email: satishogale@iiserpune.ac.in, satish.ogale@tcgcrest.org, s.patil@iiserpune.ac.in

**Author Contributions**

# Authors contributed equally. V.A., A.B. and S.O. designed and initiated the research work. A.B. carried out the thin film growth and XRD characterizations. V.A. and P.A.S. accomplished all *I-V* measurements as well as related analysis. S.R. and S.P. contributed to AFM measurements and related analysis. S.D. carried out XPS characterizations, V.A., P.A.S., A.B. and S.O. contributed to XPS analysis. S.P. and M.S. helped in interpretations and critical analysis. V.A., P.A.S., A.B., and S.O. wrote the manuscript with the input from all authors. All authors have given approval to the final version of the manuscript.

**Funding Sources**

DST Nanomission (Govt. of India), DST-SERB (Govt. of India)

ACKNOWLEDGMENT

S.O. acknowledges the funding of DST Nanomission (Thematic unit, SR/NM/TP13/2016), India. V.A. is grateful to DST-SERB, India for the research grant under TARE-2018 scheme (SERB File No.: TAR/2018/000807). P.A.S. is thankful to DST-SERB, India for research grant under TARE-2019 scheme (SERB File No.: TAR/2019/000106) and DST-FIST, India (SR/FST/College-328/2016). V.A. and P.A.S. are thankful to Dr. Vishal Thakare for insightful scientific discussions on oxide ReRAMs.

# Supporting Information

**Resistive switching in $HfO_{2-x}/La_{0.67}Sr_{0.33}MnO_3$ heterostructure: An intriguing case of low H-field susceptibility of an E-field controlled active interface**

*Vivek Antad,*[1,2,#]*, Parvez A. Shaikh,*[3,#] *Abhijit Biswas,*[4,#] *Shatruhan Rajput,*[1] *Shrinivas Deo,*[5] *ManjushaShelke,*[6] *Shivaprasad Patil,*[1,*] *and Satishchandra Ogale*[1,7,*]

[1]Department of Physics and Centre for Energy Science, Indian Institute of Science Education and Research (IISER) Pune, Pune 411008, India

[2]Department of Physics, MES's Nowrosjee Wadia College of Arts and Science, Pune 411001, India

[3]Department of Physics, Y & M AKI's Poona College of Arts, Science and Commerce, Pune 411001, India

[4]Department of Materials Science and Nanoengineering, Rice University, Houston, Texas 77005, USA

[5]Centre for Materials Characterization, CSIR-NCL, Pune 411008, India

[6]Physical and Materials Chemistry Division, Polymer and Advanced Materials Laboratory, CSIR-NCL, Pune 411008, India

[7]Research Institute for Sustainable Energy (RISE), TCG Centres for Research and Education in Science and Technology (TCG-CREST), Kolkata 700091, India

***Corresponding author:**

S.O. Email: satishogale@iiserpune.ac.in; Tel: +91 (20) 2590 8292

S. P. Email: s.patil@iiserpune.ac.in; Tel: +91 (20) 2590 8034

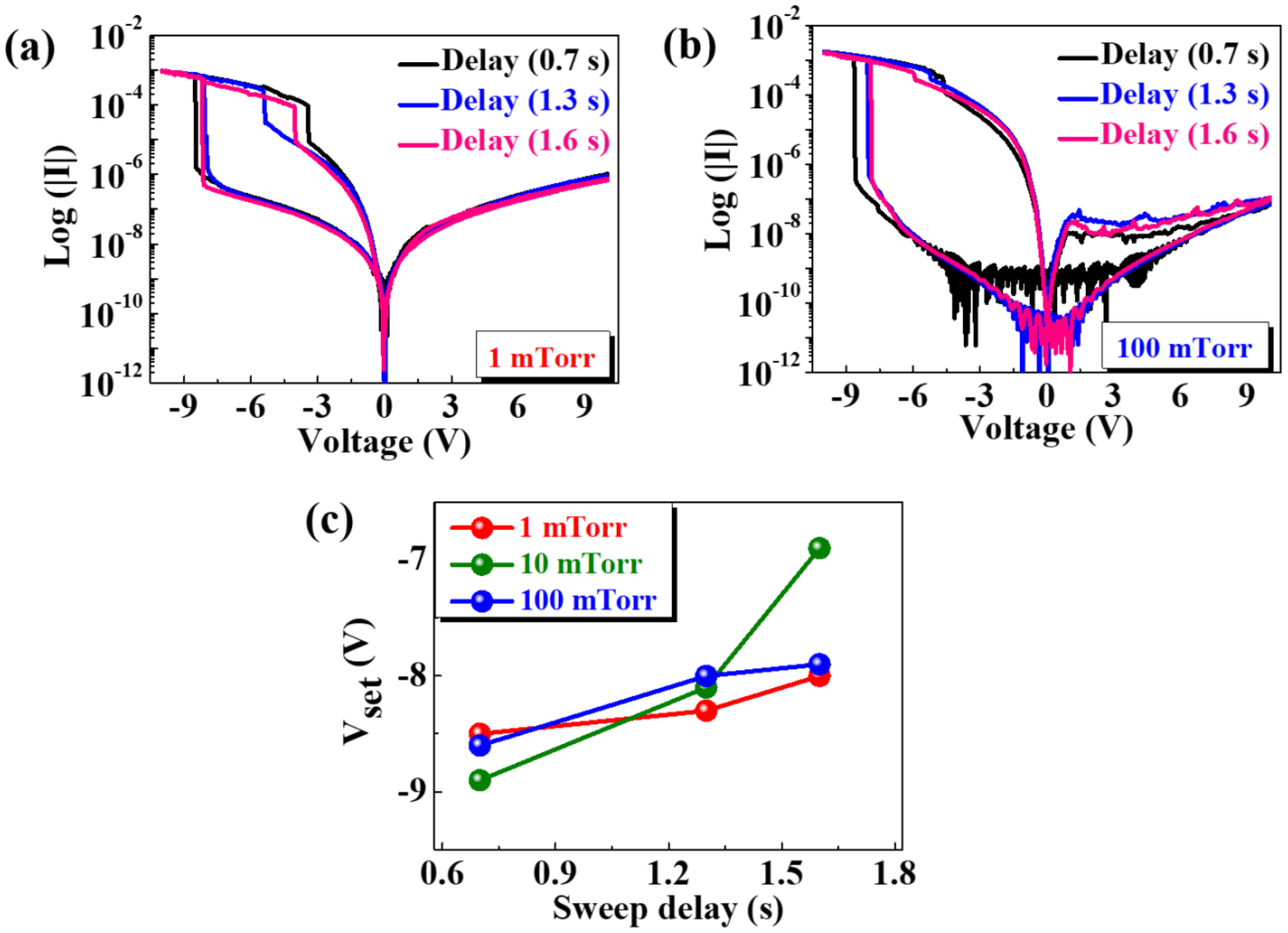

**Figure S1.** Influence of voltage sweep delays (0.7s, 1.3s, and 1.6s) on RS of $HfO_2$/LSMO/LAO devices where $HfO_2$ was grown at: (a) 1 mTorr and (b) 100 mTorr $O_2$ pressures. (c) $V_{set}$ as a function of sweep delay for $HfO_2$/LSMO devices, with $HfO_2$ grown at various $O_2$ pressures.

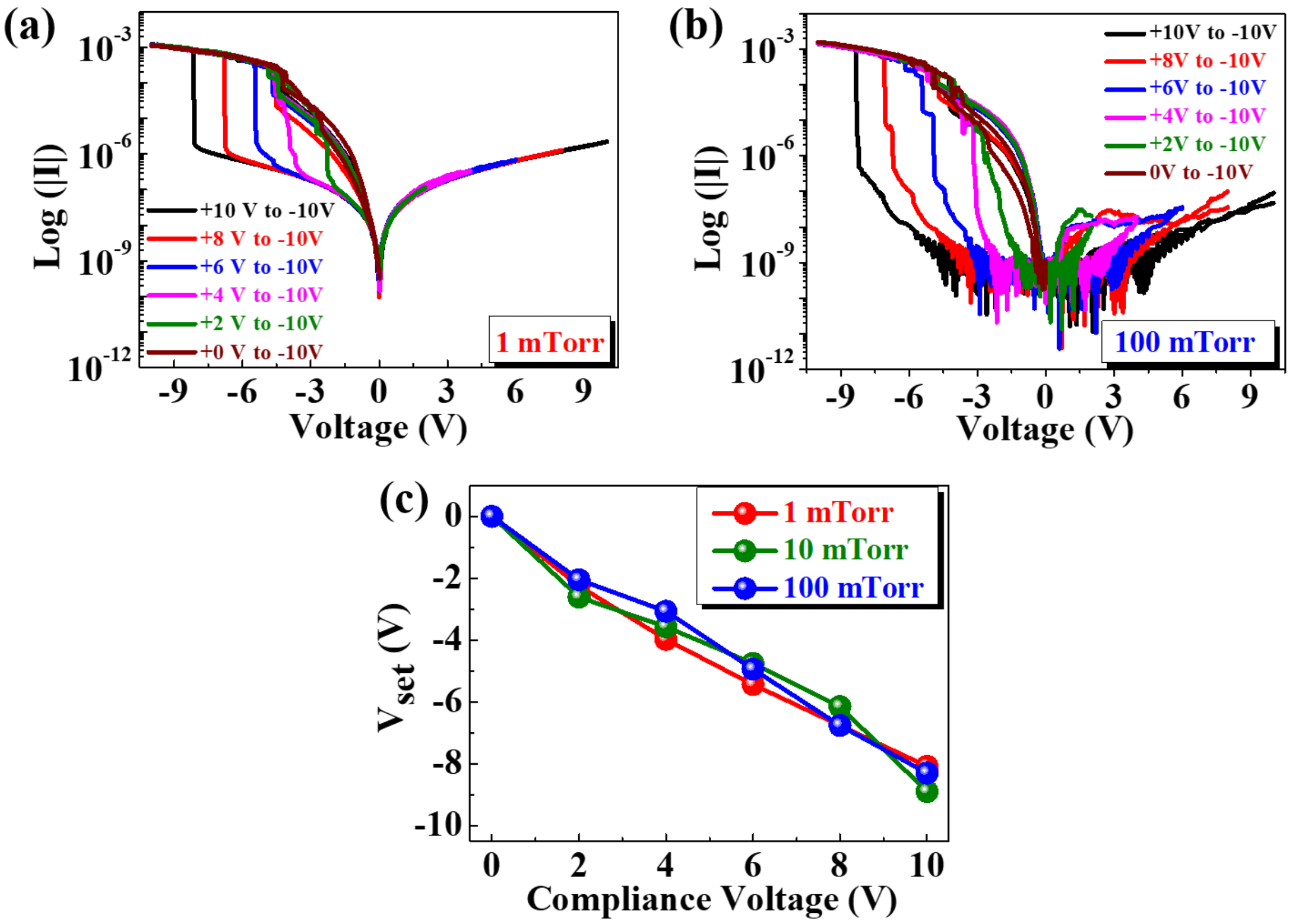

**Figure S2.** Influence of positive compliance voltages on RS of $HfO_2$/LSMO/LAO devices where $HfO_2$ was grown at: (a) 1 mTorr and (b) 100 mTorr $O_2$ pressures. (c) $V_{set}$ vs. applied positive compliance voltages for the $HfO_2$/LSMO devices, with $HfO_2$ grown at various $O_2$ pressures.

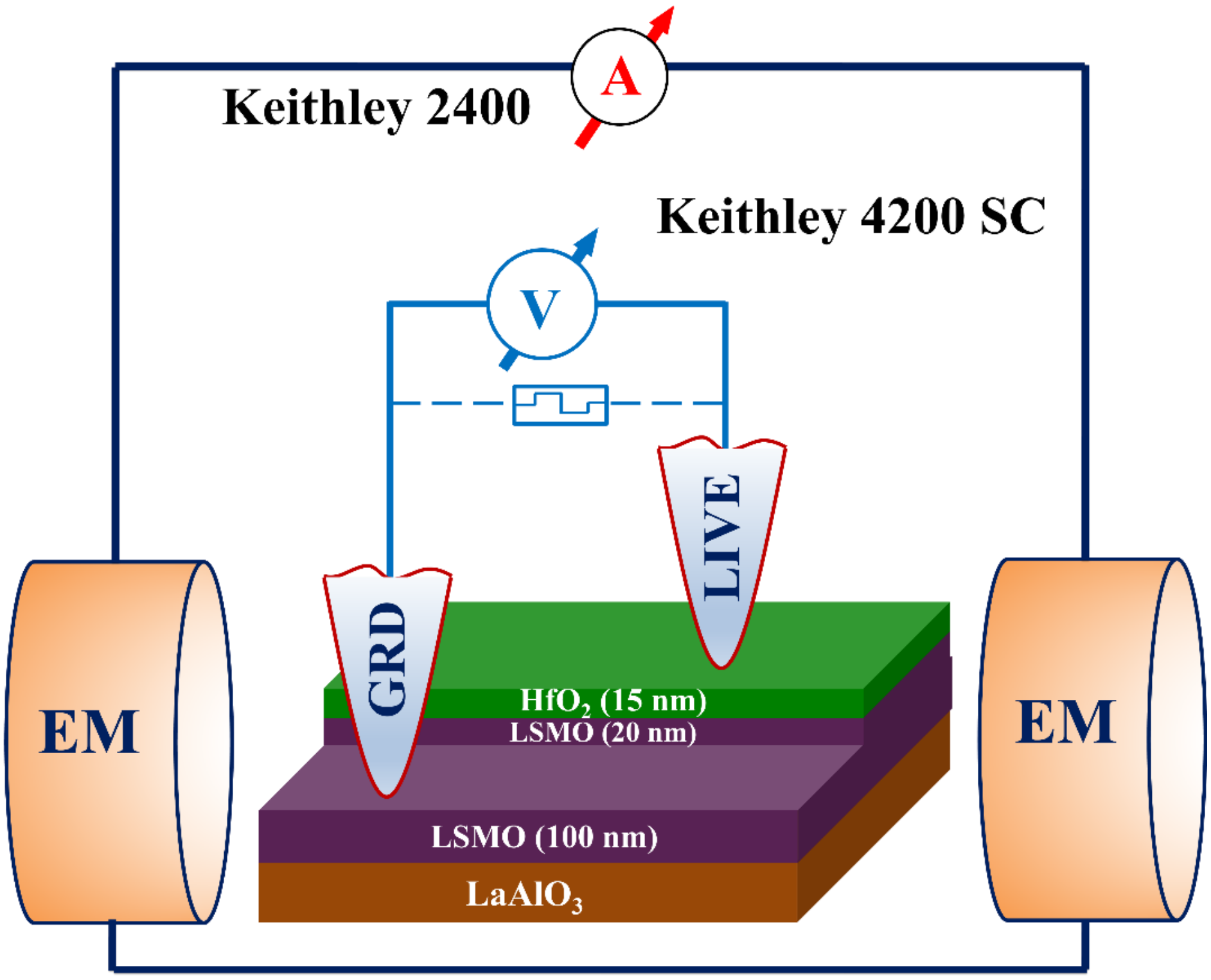

**Figure S3.** Schematics of electromagnets mounted around the RS device for the application of external magnetic field.

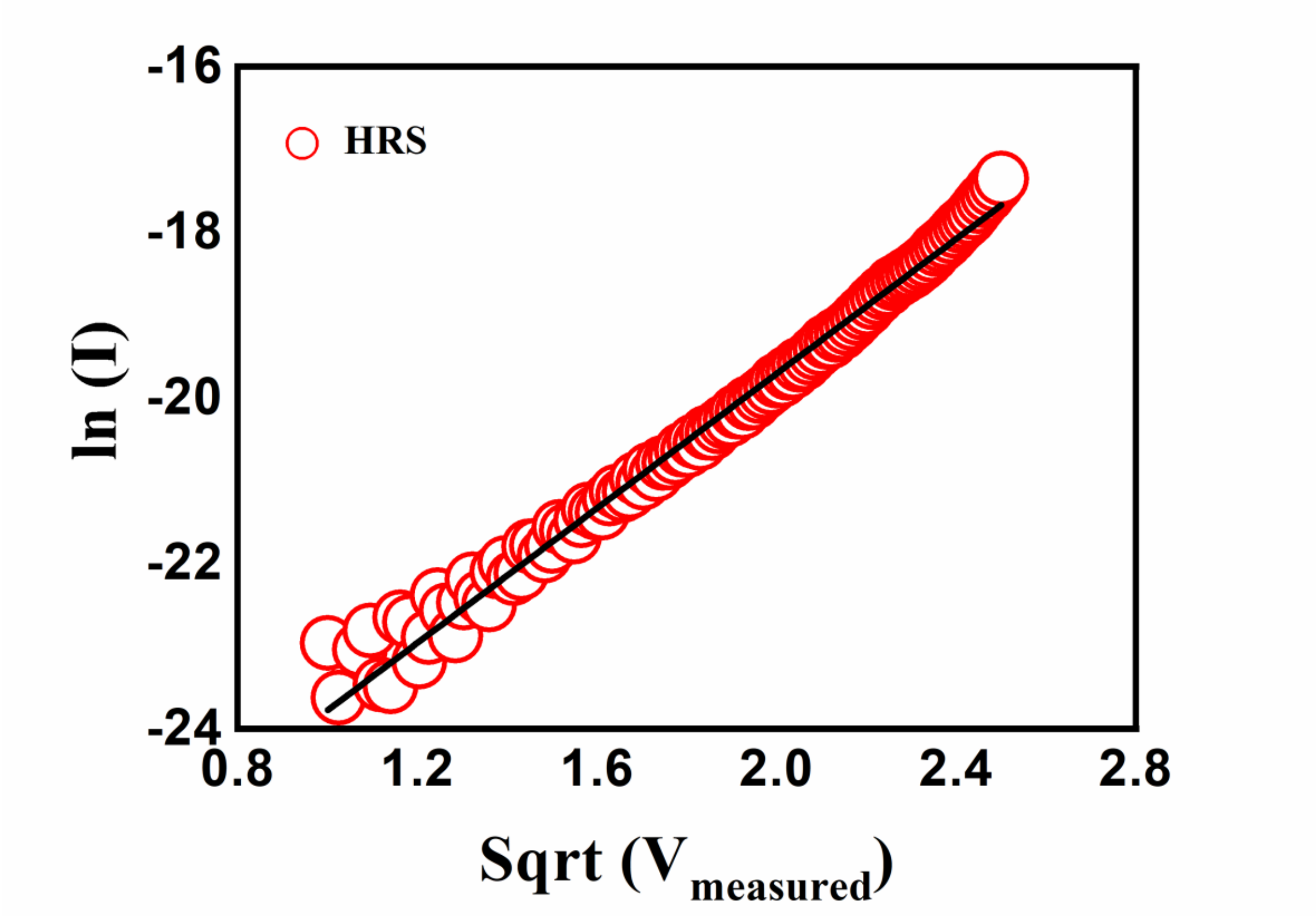

**Figure S4.** ln $(I)$ $Vs.$ $\sqrt{V}$ plot showing the linear dependence in the negative voltage region for HRS condition of the RS device.

**Supporting Information S5**

## X-ray Photoelectron Spectroscopy of the $HfO_2$/LSMO Heterostructure.

In support of the arguments made regarding the mechanism we contemplated the use of XPS technique to elucidate the valence of Mn in the fresh sample and the sample subjected to the electric field. However, this issue is non-trivial because here we are addressing a buried interface (LSMO/$HfO_2$ interface below 10-15 nm over-layer thickness of $HfO_2$) and the valence changes happening therein after application of electric fields in the Resistive Switching (RS) action. The XPS technique (accessible to us for ascertaining valence states) can only access the top 5-10 nm, as is well known. Thus, one has to sputter out much of the top layer and then look at the interface. This is destructive and can also potentially change the valence states due to ionic diffusion(s). Particularly, for heavy element like Hf the sputtering requires high energy of ions and this can induce rapid oxygen diffusion changing valence picture entirely. Non-invasive depth profiling techniques involving synchrotron could be somewhat useful (though not highly depth sensitive for what one may be looking at here), but these are not easily accessible and would require several months of separate effort.

Therefore, we decided to work near the step-edge where the thickness of the top layer could wane out a bit gradually and the under-layers could give some desired signal. This indeed gave us a fairly good picture, as shown by the results below, supporting the arguments made in the paper about the mechanism.

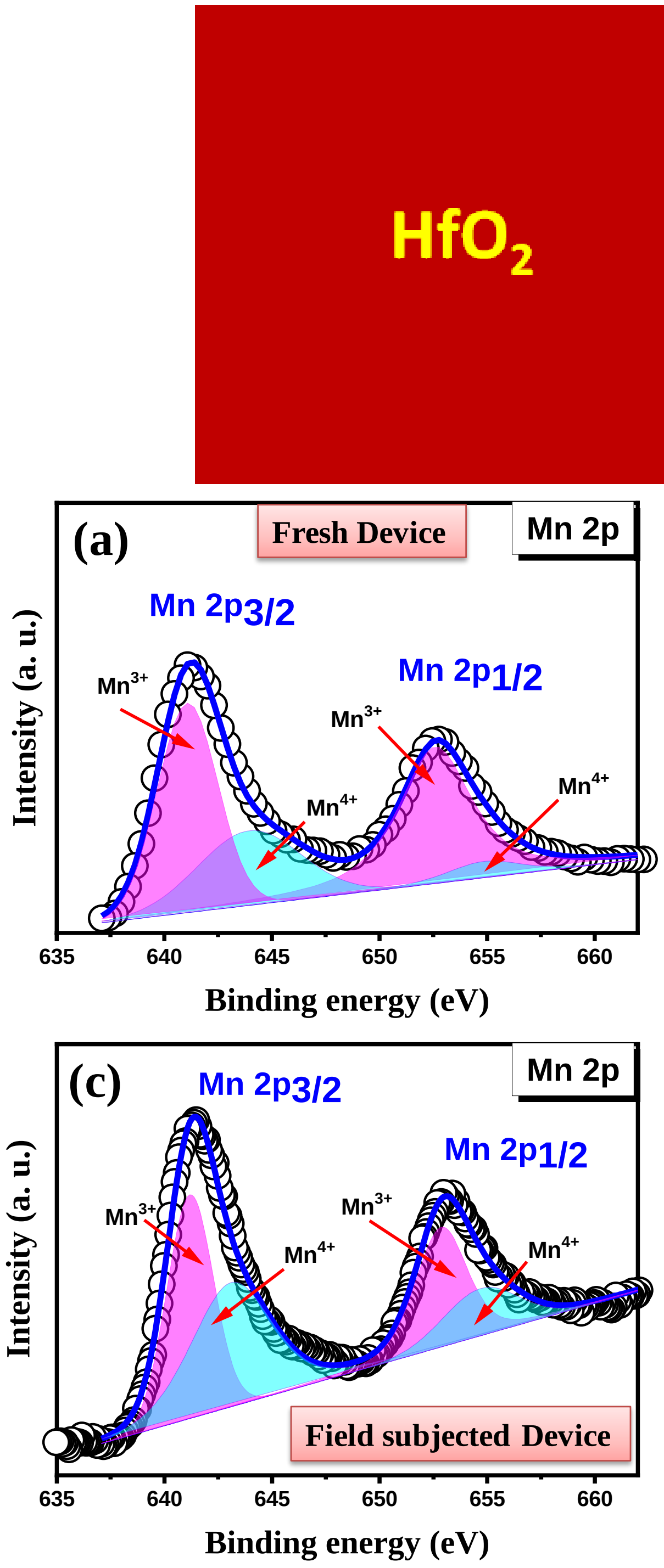

**Figure S5 (I):** Mn 2p core and Hf 4f core XPS of $HfO_2$/LSMO devices (fresh and field subjected) at the step-edge (black rectangle in schematics) showing the presence of and changes in the mixed valence states of both layers.

In order to justify our point about the issues of destructive testing by sputtering, we also carried out depth profiling by sputtering of the top $HfO_2$ layer and then recording XPS for both Hf and Mn. As can be seen from the Figure below, progressively the Hf signal decreases while Mn signal increases, as expected.

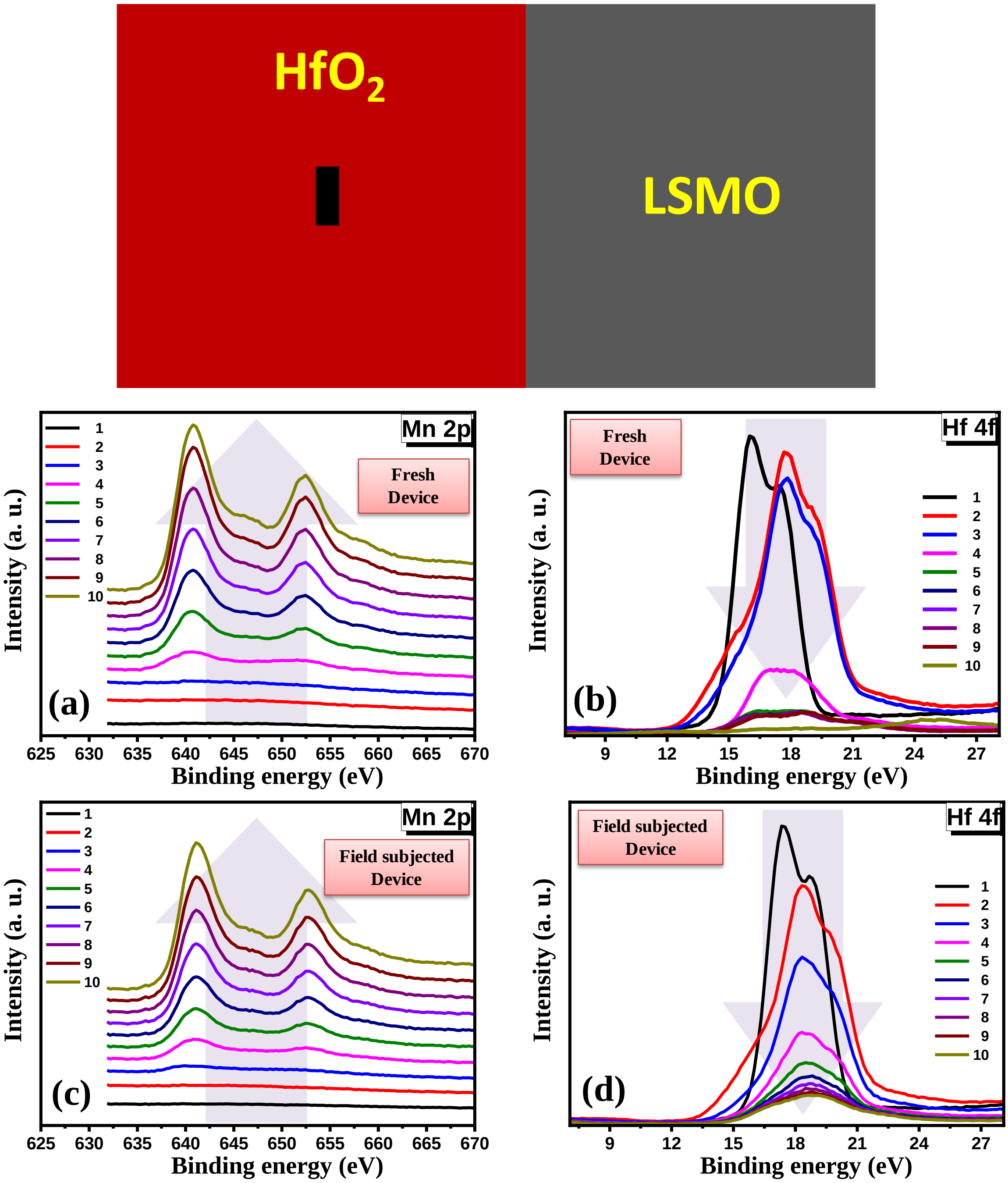

**Figure S5(II):** Mn 2p core and Hf 4f core XPS of $HfO_2$/LSMO devices (fresh and field subjected) with depth profiling scan captured after each sputtering step.

In the following Figure we show fitting of the data for an intermediate case for which both the Mn and Hf signals are clear.

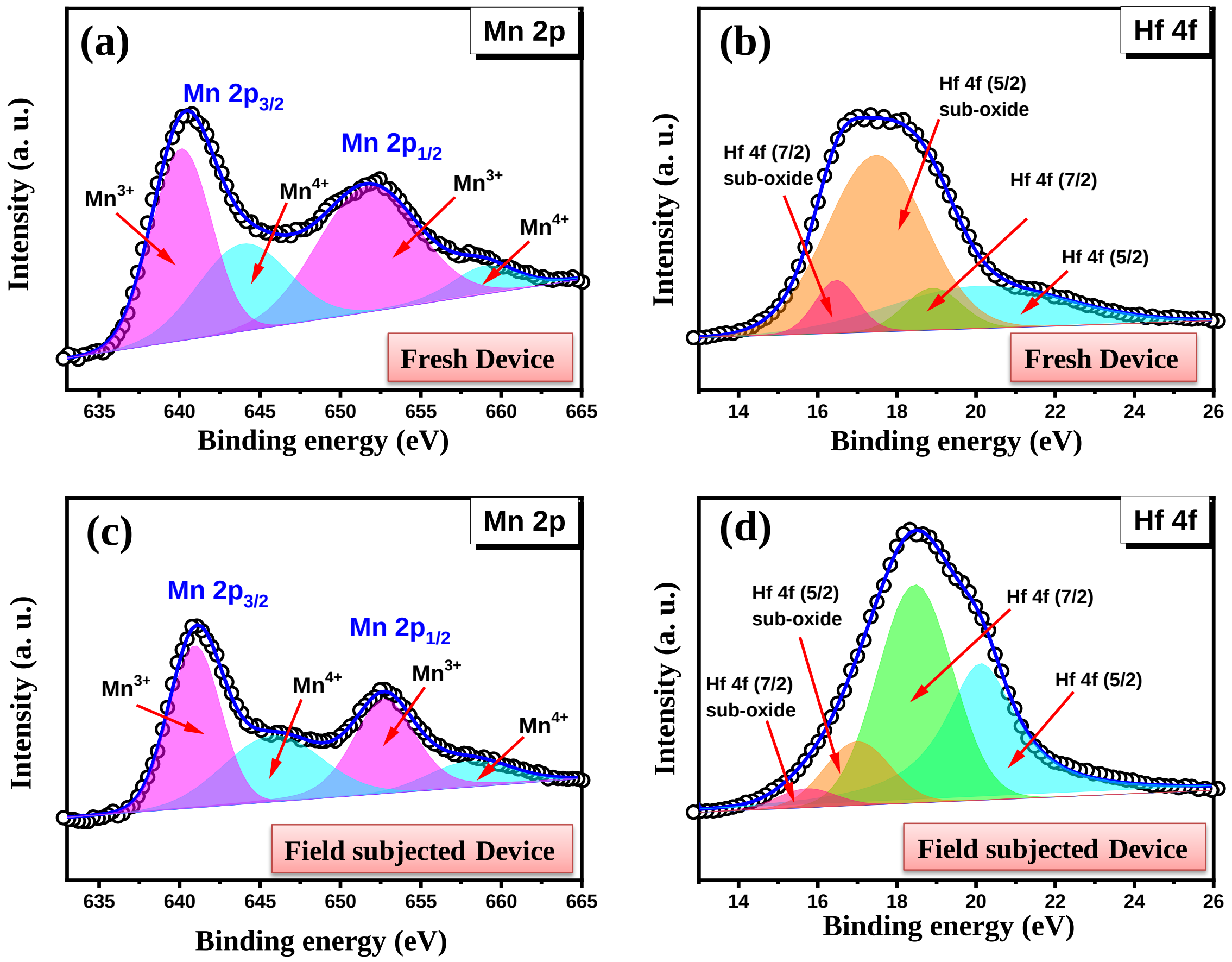

**Figure S5(III):** Mn 2p core and Hf 4f core XPS of $HfO_2$/LSMO devices (fresh and field subjected) after depth profiling showing the presence of and changes in the mixed valence states of both layers.

It can be seen that the valence states are significantly affected when compared with the un-sputtered case data recorded at the step edge. This proves that for the case of RS buried interface destructive technique is not proper and could lead to improper conclusion. Hence our analysis of data near the step edge is the one which is the most reasonable.

We have mentioned the above matters in the paper briefly and presented the new XPS data in the supporting information.

**Table S1.** Magnetic field effect on the resistive switching parameters of devices comprised of different material systems, compared with the current research work.

| Article | Journal | Magnetic Field | Material-system | Change in switching voltage | Change in switching order | Reasoning behind observed changes |
|---|---|---|---|---|---|---|
| **Antad *et al.* (Current work)** | ---- | **0 mT to 30 mT (Parallel to device)** | **$HfO_2$ (15 nm) /LSMO(20 nm) /LAO** | **increases from -9V to -8V (for 10 mT)** | **~$10^5$ to $10^0$ order (At 30 mT)** | **Redistribution of $Mn^{3+}$ and $Mn^{4+}$ ion concentrations and trapping/de-trapping of electrons at $HfO_2$/LSMO interface due to change in LSMO conductance in the presence of magnetic field** |
| Sahu *et al.*[1] | Nature Sci. Rep. | 36 mT to 300 mT (Parallel to device) | Ag/$TiO_2$ (5 $\mu$m)/FTO | increases from +4V to +4.4V | $10^{1/2}$ to $10^1$ | Residual Lorentz force |
| Fang *et al.*[2] | Physica B | 550 mT (Perpendicular to device) | $SrTiO_3$:Nb/ZnO/Au (Film thickness not mentioned) | -- | No RS to $10^1$ (for 60 min exposure) | Increase in the density of interfacial states responsible for trapping/de-trapping of electrons |
| Jena *et al.*[3] | APL | 25 mT to 333 mT (Parallel to device) | Ag/$BiFeO_3$/FTO (BFO thickness not mentioned) | increases from +3V to +4V | Order remains almost same | Lorentz force and magneto-electric effect |
| Li *et al.*[4] | Phy. Lett. A | 100 mT to 800 mT (Parallel to device) | $BaTiO_3$ (50nm) /FeMn (20 nm) /$BaTiO_3$ (50 nm) | decreases from +2.8V to +1.7 V | $10^1$ to $10^2$ | Ohmic and space charge limited conduction mechanism, and redistribution of bound magnetic polaron |
| Sun *et al.*[5] | RSC Adv. | 1500 mT and 2500 mT (Parallel to device) | Ag/[$BiFeO_3$/γ-$Fe_2O_3$] (5 $\mu$m)/FTO | 0.82V to +0.92V | $10^3$ to $10^2$ Order decreases | Improvisation in the ferroelectric polarization of $BiFeO_3$ due to strong magneto-electric effect, increase of the extra opposite-direction electrical field, generated by the applied magnetic field. |
| Wang *et al.*[6] | APL | 0 mT to 2000 mT (Perpendicular to device) | Ag (20 nm)/ Mg (60nm)/ MgO (20nm)/ $SiO_2$-Si | from 7V to 8.3V | Order decreases from $10^1$ to 1.1 | Increase in the switching voltage attributed to modification in the Laudau levels of electrons and the thermal ionization after generation and recombination processes, while suppression of RS explained on the basis of MR generated at MgO/SiO2 interfaces when magnetic field reaches to 2000 mT |

## Table of contents

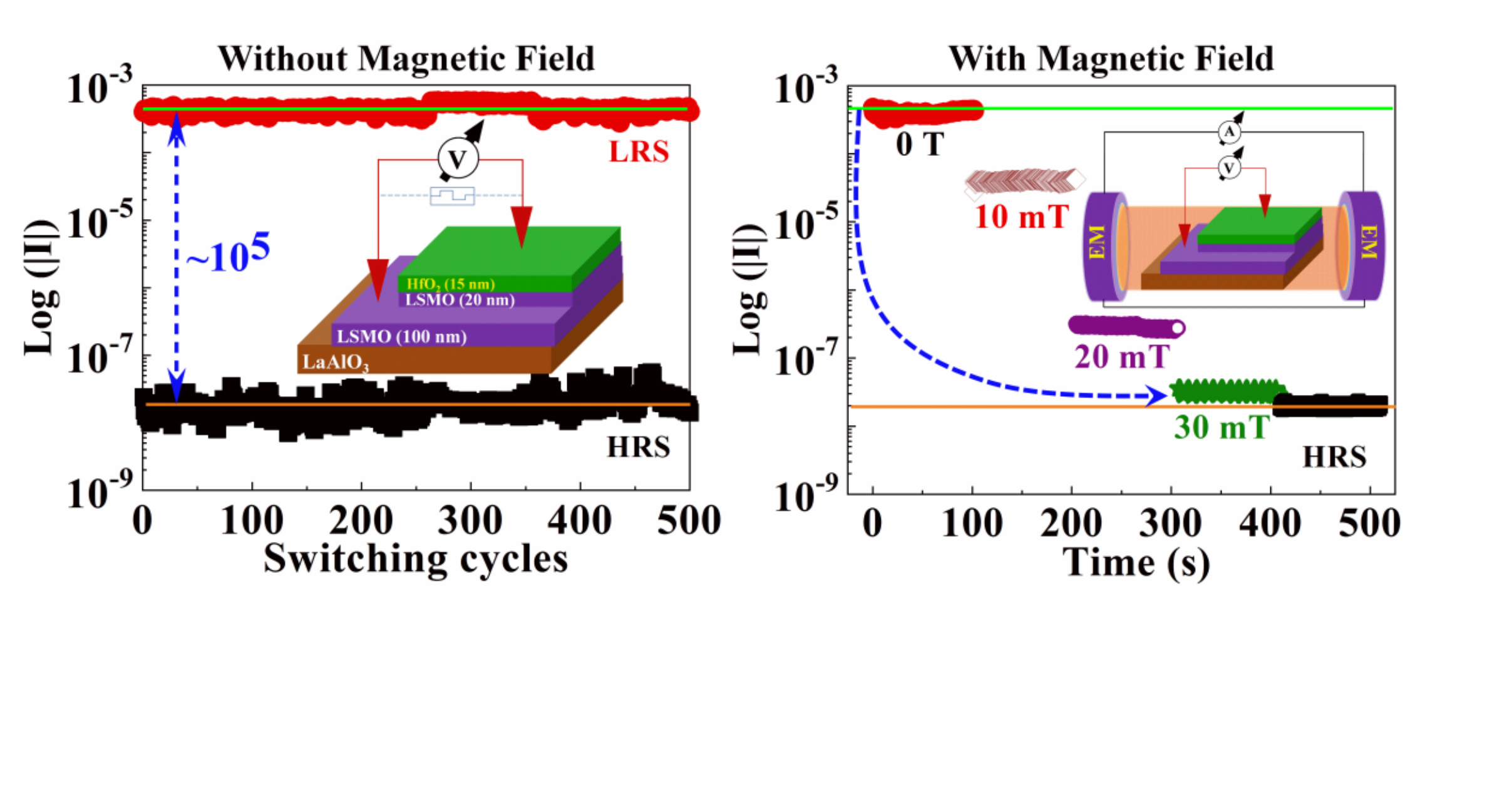